%% file: 00-paper.tex
\documentclass{article}

\newcommand{\card}[1]{\lvert #1 \rvert} 

\usepackage{ijcai19}

\usepackage{amsmath}

\usepackage{rotating}
\usepackage{xcolor}
\usepackage{calc, enumitem} 
\usepackage{amssymb}
\usepackage{multirow}
\usepackage{color, colortbl}
\usepackage{array}

\usepackage{times}
\usepackage{soul}
\usepackage[hyphens]{url}
\usepackage[hidelinks]{hyperref} 
\usepackage[utf8]{inputenc}
\usepackage[small]{caption}
\usepackage{graphicx}
\usepackage{amsmath}
\usepackage{booktabs}
\definecolor{ControlColor}{RGB}{84,130,53}
\definecolor{SaliencyColor}{RGB}{253,192,134}
\definecolor{RewardsColor}{RGB}{255,255,153}
\definecolor{EverythingColor}{RGB}{56,108,176}

\definecolor{highlightColor}{RGB}{255,255,153} 
\include{z-commands}

\input{000-titleAuthorAbstract.tex}

\begin{document}

\maketitle

\begin{abstract}
We present a user study to investigate the impact of explanations on non-experts' understanding of reinforcement learning (RL) agents.  We investigate both a common RL visualization, saliency maps (the focus of attention), and a more recent explanation type, reward-decomposition bars (predictions of future types of rewards). We designed a 124 participant, four-treatment experiment to compare participants' mental models of an RL agent in a simple Real-Time Strategy (RTS) game.  Our results show that the combination of both saliency and reward bars were needed to achieve a statistically significant improvement in mental model score over the control. In addition, our qualitative analysis of the data reveals a number of  effects for further study.  
\end{abstract}


\input{1-Introduction.tex}
\input{2-background.tex}
\input{3-Methodology.tex}
\input{4-MentalModel.tex}
\input{5-PredictionResults.tex}
\input{7-ThreatsToValidity.tex}
\input{8-conclusion.tex}\clearpage

\bibliographystyle{named}
\bibliography{00-paper}

\end{document}

%% file: z-commands.tex
\newcommand{\briefCite}[1]{~\citeauthor{#1} (\citeyear{#1})}

\newcommand{\mysection}[1]{\section{#1}}
\newcommand{\mysubsection}[1]{\subsection{#1}}

\usepackage{adjustbox}
\newcommand{\colorboxBackgroundForegroundText}[3]
{\protect\adjustbox{padding=1pt 1pt, bgcolor=#1}{\color{#2}#3}}

\newcommand{\user}[3]{{%
~(#1#2)}}

\newcommand{\anova}[3]{(ANOVA, F = #1, df = (1,#2), p = #3)}

\newcommand{\mapsize}{1.64cm}

\newcommand{\redact}[1]{X University}

\newcommand{\treatment}[1]{{%
#1}}

\newcommand{\feature}[1]{{%
#1}}

\newcounter{fixmeSectionCounter}
\newcounter{fixmeTotalCounter}

\makeatletter
\@addtoreset{fixmeSectionCounter}{section}
\@addtoreset{fixmeSectionCounter}{subsection}
\@addtoreset{boldifyCounter}{section}
\@addtoreset{boldifyCounter}{subsection}
\makeatother

\newcommand{\boldify}[1]{}
\ifdefined\boldifyON
	\renewcommand{\boldify}[1]{\par\noindent%
		\textbf{{\color{green}**}%
		~\arabic{section}.\arabic{subsection}%
		: #1\\}
	}
\fi

\newcommand{\fakeResult}[1]{}
\ifdefined\fakeResultON
    \renewcommand{\fakeResult}[1]{\par\noindent%
        \textbf{{\color{blue}**FAKE FAKE FAKE**}%
        ~\arabic{section}.\arabic{subsection}%
        : #1\\}
    }
\fi

\newcommand{\FIXME}[1]{}
\ifdefined\fixmeON
	\renewcommand{\FIXME}[1]{\par\noindent%
		\stepcounter{fixmeSectionCounter}\stepcounter{fixmeTotalCounter}%
		{\color{red}\fbox{\color{black}%
			\parbox{.97\linewidth}{%
				\textbf{FIXME \arabic{section}.\arabic{subsection}.%
        		\arabic{fixmeSectionCounter} (\color{red}%
        		\#\arabic{fixmeSectionCounter}):} #1}
        		}%
        }
	}
\fi

\ifdefined\specialFooterON
\usepackage{fancyhdr, datetime}
\fi

%% file: 000-titleAuthorAbstract.tex
\title{Explaining Reinforcement Learning to Mere Mortals: An Empirical Study}

\author{Andrew Anderson \and Jonathan Dodge\and Amrita Sadarangani \and Zoe Juozapaitis \and  \\Evan Newman \and Jed Irvine \and Souti Chattopadhyay \and Alan Fern \normalfont and \textbf{Margaret Burnett}
\affiliations
Oregon State University
\emails
\{anderan2, dodgej, sadarana, juozapaz, newmanev, irvine, chattops, Alan.Fern, burnett\}@eecs.oregonstate.edu}

%% file: 1-Introduction.tex
\section{Introduction}

\boldify{XAI is important, so are soundness and completeness in explanation.
Consequence of bad explanation is a bad mental model}

Although eXplainable Artificial Intelligence (XAI) has seen increasing interest as AI becomes more pervasive in society,  
much of XAI work does not attend to the \emph{people} who consume explanations.
In this paper, we draw upon a work that does, which introduced 4 principles for explaining AI systems to people who are not AI experts~\cite{kulesza2015principles}.
These principles were: be iterative, be sound, be complete, and do not overwhelm the user, where here the notions of soundness and completeness are analogous to ``the whole truth (completeness) and nothing but the truth (soundness).''
We ensured that our explanations were ``sound'':
we did not approximate/simplify them.
They were also ``complete,'': every agent input \& output was represented in the UI.
\FIXME{JED: Improve intuition for soundness/completeness with text from rebuttal  (R3-10)}

\boldify{And they make MMs better.  So what is a mental model and what does it have to do with XAI?}

Empirical results showed that explanations adhering to these principles enabled non-AI experts to build higher-fidelity mental models of the agent than non-AI experts who received less sound/complete explanations~\cite{kulesza2015principles}.
People's mental models, in the context of XAI, are basically their understanding of the way the agent works. 
More formally, mental models are, ``\emph{internal representations that people build based on their experiences in the real world.}''~\cite{norman1983mental}.
People's mental models vary in complexity and accuracy, but
a good mental model will enable a person to \emph{understand} system behavior, and a very good one will enable them to \emph{predict} future behaviors.

\boldify{How did we effect a change in participants' mental models? Explanations, namely saliency and reward}

In this paper, we investigate how people's mental models of a reinforcement-learning agent vary in response to different visual explanation styles--saliency maps showing where the agent is ``looking,'' and reward decomposition bars showing the agent's current prediction of its future score.
To do so, we conducted a controlled lab study with 124 participants across four treatments (saliency, rewards, both, neither), and measured both their understanding of the agent and their ability to predict its decisions.
Our investigation was in the context of Real-Time Strategy (RTS) games.

\FIXME{JED: If we are going to justify our choice of explanation types, we do it here. (R3-9) }

\boldify{Mental models of the agent and the domain it operates in are interrelated, so DESCRIBE the domain (Starcraft, which has some formative studies we can stand on).}

However, publicly available RTS games have stringent time constraints, complex concepts, and myriad decisions, which would have introduced too many confounding variables for a controlled study.
For example, we needed each participant to consider the \emph{same} set of decisions.
Thus, we built our own game, inspired by RTS, which we describe later.

In this context, we structured our investigation around the following research questions:

\begin{itemize}[topsep=0pt,itemsep=0pt,partopsep=0pt, parsep=0pt]
    \item \textbf{RQ-Describe} - Which treatment is better (and how) at enabling people to \emph{describe} how the system works?
    \item \textbf{RQ-Predict} - Which treatment is better (and how) at enabling people to \emph{predict} what the system will do?
\end{itemize}


%% file: 2-background.tex
\section{Background \& Related Work}



\boldify{Traditional RL framework uses a scalar-valued reward. This is not a rich representation, and is essentially limited to taking maximums and such.}

We focus on model-free RL agents that learn a Q-function $Q(s,a)$ to estimate the expected future cumulative reward of taking action $a$ in state $s$. 
After learning, the agent greedily selects actions according to $Q$, i.e. selecting action $\arg\max_a Q(s,a)$ in $s$.
RL agents are typically trained with scalar rewards, leading to scalar Q-values.  
While a human can compare the scalars to see how much the agent prefers one action over another, the scalars give no insight into the cost/benefit factors contributing to action preferences. 

\boldify{We can use Reward Decomposition to form a vector-valued!
Now we can compare actions in semantically meaningful ways (e.g. this action is predicted to result in the most enemies destroyed)}

To address this problem, we draw on work by~\cite{erwig2018explaining} for reward decomposition, which exploited the fact that rewards can typically be grouped into semantically meaningful types.
For example, in RTS games, reward types might be ``enemy damage'' (positive reward) or ``ally damage'' (negative reward).
Reward decomposition exposes reward types to an RL agent by specifying a set of types $C$ and letting the agent observe, at each step, a $\card{C}$-dimensional decomposed reward vector $\vec{R}(s,a)$, which gives the reward for each type. The total scalar reward is the sum across types, i.e. $R(s,a)=\sum_{c\in C} \vec{R}_c(s,a)$. The learning objective is still to maximize the long-term scalar reward. 


By leveraging the extra type information in $\vec{R}(s,a)$, the RL agent can learn a decomposed Q-function $\vec{Q}(s,a)$, where each component $\vec{Q}_c(s,a)$ is a Q-value that only accounts for rewards of type $c$. 
Using the definition of $R(s,a)$, the overall scalar Q-function can be shown to be the sum of the component Q-functions, i.e. $Q(s,a)=\sum_{c} \vec{Q}_c(s,a)$. 
Prior work has shown how to learn $\vec{Q}(s,a)$ via a decomposed SARSA algorithm \cite{russell2003q,erwig2018explaining}.


Before~\briefCite{erwig2018explaining}, others considered using reward decomposition~\cite{russell2003q,van2017hybrid}---but for speeding up learning. 
Our focus here is on their visual explanation value.  
For a state $s$ of interest, the decomposed Q-function $\vec{Q}(s,a)$ can be visualized for each action as a set of ``reward bars'', one bar for each component.  
By comparing the bars of two actions, a human can gain insight into the trade-offs responsible for the agent's preference. 

\boldify{Alan: The last sentence of the previous paragraph could be dropped if needed, but better to keep if we can to avoid complaints about not citing everything that used reward decomposition.}



\boldify{Related work has also used reward decomposition for explanation, but we focus on non-experts and do not use MSX.}

\boldify{Alan note. I don't think a discussion of the MSX is needed. We don't use it and there is no good reason to explain why we don't.}



\boldify{Saliency maps are a visual explanation, used to pinpoint the exact pixels/neurons that a neural net paid attention to while classifying}

\boldify{Alan Note: I don't think we need to go into the details and disadvantages of the saliency methods. The chosen set is a bit arbitrary and we are not trying to make a point of which saliency method is best. I've tried to adjust so that we say what saliency is, say there are many methods, and then say what we use.}


\boldify{Alan: I removed reference to RTS game details below, since those will be covered later.}


Instead of the rewards, a human may want to know which parts of the agent's input were most important to the value computed for a reward bar (i.e. a particular $\vec{Q}_c(s,a)$). 
Such information is often visualized via saliency maps over the input.
Our agent uses neural networks to represent the component Q-functions, letting us draw on the many neural network saliency techniques (e.g. \cite{simonyan2013deep,springenberg2014striving,zeiler2014visualizing,zhang2018top}). While there have been a number of comparison and utility studies (e.g.~\cite{adebayo2018Sanity,ancona2018towards,greydanus18a,kim2018TCAV,riche2013saliency}), there is no consensus on a best way.

After exploring various techniques, we modified \briefCite{fong2017interpretable}'s work on image classification, using a perturbation method that focused on \emph{attributes} (blocks of pixels), rather than individual \emph{pixels}, as used in computer vision, to aid human interpretation.
Since the network may ``focus'' on different parts of the input for each reward bar, we compute saliency maps for each one, which the UI could visualize.

\FIXME{AAA @JED: (R2-2)}

\boldify{``Traditional'' Saliency suffers from a limitation uses weights only, not bias or inputs (it was not clear to me if this refers to gradient-based techniques or what. Adjust this boldify as you incorporate the fix, and delete the previous boldify if I have captured the goal of this paragraph)}
\boldify{Alan note: the above is not important.}


\boldify{Perturbative saliency overcomes these limitations, so we used it with some adaptation.}
\boldify{Alan Note: I adjusted the following so that is just states what we use. I'm going to ignore some of the details here, since it would get way too technical and is not really important.}

\boldify{Saliency is assumed to be useful in visual explanations, but has not really been confirmed as such.
First, there are many flavors of saliency, some of which have already been demonstrated to be insufficient.
It has been evaluated with humans several times with generally inconclusive results.}
\boldify{Alan note: I moved this above.}


%% file: 3-Methodology.tex
\section{Methodology}

\boldify{Here is the 10,000 foot view of our study, being clear about Independent Variable (IV) and Dependent Variable (DV) and how we measured it}

 We performed an in-lab study using a between-subjects design with \textbf{explanation style} (Control, Saliency, Rewards, Everything) as the  independent variable.
Our dependent variable was the quality of participants' mental models--measured by analysis of two main data sources: 
1) answer to a post-task question,
2) accuracy of participants' prediction for the agent's selected action at each decision point (DP).

We ran an ablation study, where we measured the impact of each explanation by adding or removing them, as shown in Figure~\ref{fig:userInterface}.
Thus, Everything - Rewards - Saliency = Control, as follows:
\textbf{1.} \treatment{Control} participants saw only the agent's actions, its consequences on the game state and the score (Region~1 \& Figure~\ref{fig:sequentialDecisions}), and question area (Region~4).
\textbf{2.} \treatment{Saliency} participants saw Regions~1 \& 4 plus Region~2, allowing them to infer intention from gaze~\cite{newn2016exploring}.
\textbf{3.} \treatment{Rewards} participants saw Region~1 \& 4 plus Region~3, allowing them to see the agent's cost/benefit analysis.
\textbf{4.} \treatment{Everything} participants saw all regions.


\subsection{Participants And Procedures}

\boldify{The participants were....}

\FIXME{AAA @ all - R2-3
JED: This is in response to R2-3, but there was also a question about recruitment mechanism. Suggest:

After our study was approved by 2 ethics committees, we recruited via flyers linked to an online survey at Oregon State University.
208 people responded, of which 124 participants completed the study}

With 2 ethics committees' approval, we ran 124 participants from 208 online survey respondents at Oregon State University.
Since we were interested in AI non-experts,  our selection criteria excluded Computer Science majors and anyone who had taken an AI course.
We assigned the participants to a two-hour in-lab session based on their availability and randomly assigned a treatment to each session. 

We collected the following demographics: major and experience with RTS games (Table~\ref{table:participants}).
78\% of our participants, were ``Gamers,'' defined as those with 10+ hours RTS experience, consistent with prior research~\protect\cite{penney2018}.
Afterwards, we noticed that gamers were spread evenly across treatments, so it was unnecessary to control for this statistically (Figure~\ref{fig:RTSexperts}).

\input{figure/regionsUI.tex}
\input{tables/Participants}
\input{figure/RTSexperts.tex}
\input{figure/sequentialDecisions}

\boldify{...and here's what they DID}

\FIXME{AAA -(R3-3)} 
We began sessions with a 20-minute, hands-on tutorial to the \emph{interface/domain}, with 3 practice DPs.
Since participants were AI non-experts, we described saliency maps as, 
``\emph{...like where the eyeballs of the AI fall}'' 
and reward bars as, 
``\emph{...the AI’s prediction for the score it will receive in the future.}''
Participants had 12 minutes to complete DP1 and 8 minutes per DP for the remaining 13.

The agent died 4 times; each time was a task boundary. 
Each task had 3-4 DPs, chosen for a mixture of diversity (e.g., all objects shown at least once), and similarity (e.g., some maps had the same object types but different health).
At each DP, participants: (1) saw the game state with nothing else visible yet; (2) answered questions about the object they thought the agent would attack and why; (3) upon submitting their answer, they saw what the agent did and the explanation for their treatment (reward bars, saliency maps, both, or neither).
After all 14 DPs, they described the agent's decision making process, filled out a questionnaire, and received \$20.
\FIXME{AAA - (R3-7)
AAA @ JED:(R3-2, clarify DP vs task) and (R3-4, how were tasks chosen), soundness \& completeness}

\input{figure/GameObjectTable.tex}

\subsection{System Overview}

\boldify{RTS games are complex, so we operated on a restricted RTS}
Popularly available RTS games have an enormous action space --\briefCite{vinyalsAlphaStar} estimates $\approx10^{26}$ for StarCraft II.
With so many possibilities, it is not surprising that researchers have reported large differences in individual participants' focus, leading them to notice different decisions~\cite{dodge2018,penney2018}.
To avoid this, we built a game with a tightly controlled action space to control the entire software stack (UI, agent API, etc).
The game and study materials/code are~\href{https://ir.library.oregonstate.edu/concern/datasets/tt44ps61c}{here}\footnote{https://ir.library.oregonstate.edu/concern/datasets/tt44ps61c}
\FIXME{The above deals with reviewer 2's concern for us building our own game. (R2-1)}

\boldify{The restricted RTS featured a short series of sequential decisions with a small number of choices. Characterize the domain here}

In our game, the agent's goal was to maximize its score over each task (Figure~\ref{fig:sequentialDecisions}), subject to the following rules:
\begin{itemize}[topsep=0pt,itemsep=0pt,partopsep=0pt, parsep=0pt, leftmargin=*]
    \item Only Forts/Tanks could attack objects (Figure~\ref{fig:TableOfObjects}).
    \item At each DP, the agent \emph{had to} attack one of the quadrants.
    \item If agent/friendlies were damaged/destroyed, it lost points.
    \item If enemies were damaged/destroyed, it gained points.
    \item Once the agent killed something, it ``respawned'' on a new map, carrying over its health.
\end{itemize}


\subsubsection{The Reward Decomposition Implementation}

\boldify{JED: Here is how we specified the reward function (R), which is the first step in this process}




The agent used six reward types to learn its $\vec{Q}(s,a)$: \{Enemy Fort Damaged, Enemy Fort Destroyed, Friendly Fort Damaged, Friendly Fort Destroyed, Town/City Damaged, Town/City Destroyed\}. 
The RL agent used a neural network representation of $\vec{Q}(s,a)$. 
For each reward type $c$, there was a separate network for $Q_c(s,a)$ which took the state description as input--7, 40x40 greyscale image layers, each representing information about the state: \{Health Points (HP), enemy tank, small forts, big forts, towns, cities, and friend/enemy\}.
The overall scalar Q-values $Q(s,a)$ used for action selection were the sum of each $Q_c(s,a)$.   
The agent trained using the decomposed SARSA learning algorithm using a discount factor of 0.9, a learning rate of 0.1, with $\epsilon$-greedy exploration ($\epsilon$  decayed from 0.9 to 0.1). 
It trained for 30,000 games, at which point it demonstrated high-quality actions.


\subsubsection{The Saliency Map Implementation}

\boldify{Our neural net was fed 7 greyscale image layers, and here are their names.}

Given a state $s$, our perturbation-based approach produced a saliency map for each bar $Q_c(s,a)$ by giving data representing the true state $s$ and a perturbed state $s'$ (close to $s$), then subtracting the outputs for both states. 
Large output difference meant the system was more sensitive to the perturbed part of the state--indicating importance, which we showed with a brighter color. 
We chose to use a heated object scale, since~\briefCite{newn2017evaluating} found it to be the most understandable for their participants.
Our perturbations modify properties of \emph{objects} in the game state and thus modify  \emph{groups} of pixels, not individual pixels.


\boldify{Our perturbations are all inspired by semantically meaningful transformations (removing/transforming an object). 
Here is how each transformation gets applied}

Each of the perturbations represented a semantically meaningful operation:
\textbf{1.} Tank Perturbation. 
If a tank was present, we removed it by zeroing out its pixels in the tank layer.
\textbf{2.} Friend/Enemy Perturbation.
Transform an object from \emph{friend} to \emph{enemy} by moving the friend layer pixels to the enemy layer (and vice versa). 
\textbf{3.} Size Perturbation. 
Transform an object from \emph{big} to \emph{small} (or vice versa) by moving the pixels from one size layer to the other.
\textbf{4.} City/Fort perturbations.
Similarly transform whether an object is a City or Fort. 
\textbf{5.} HP Perturbation. Since HP is real-valued it is treated differently. 
We perturbed the object HP values by a small value, 30\%.
These operations were represented in five saliency maps: HP, Tank, Size, City/Fort, \& Friend/Enemy

\boldify{Now that we have a saliency map for A decision, we need to think about how to make them comparable across MANY decisions, so we normalize}

To make the saliency maps comparable across types, we found the maximum saliency value in each map for \emph{each} reward type \& class from 16,855 episodes. 
Normalizing each map by this value placed the pixel value $\in [0,1]$.

\boldify{Now that we have a 0-1 real valued saliency, we can present it many ways. 
We chose colorizing via heated object cuz of related work}


%% file: figure/regionsUI.tex
\begin{figure}
	\centering
	\includegraphics[width = \linewidth]{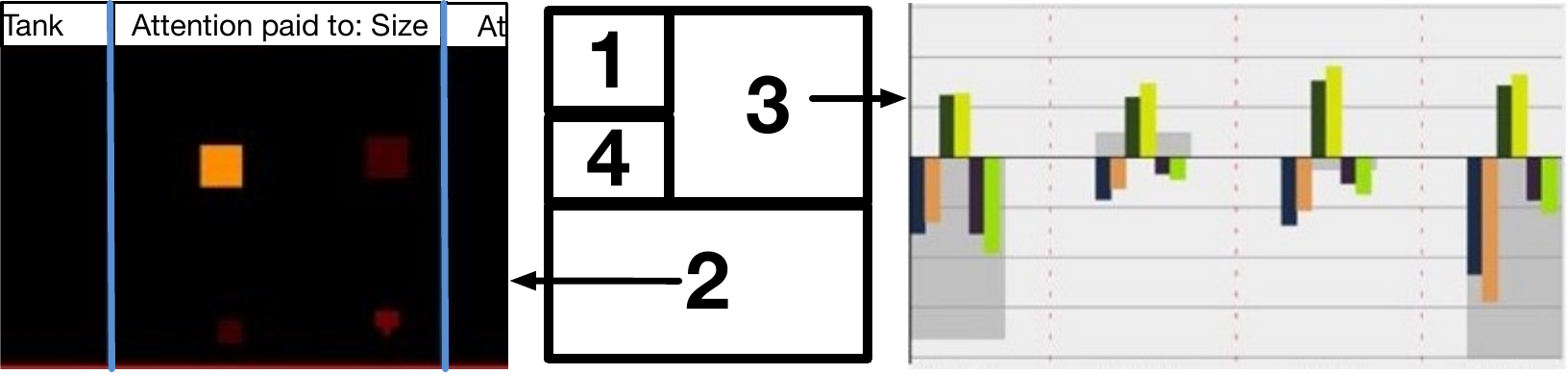}

	\caption{
    The regions of the interface.
    Region 1: game map, which we expand on in Figure~\ref{fig:sequentialDecisions}. 
    Region 2: saliency maps. 
    Region 3: reward decomposition bars for each action. 
    Region 4: participant question/response area.
    See text for who saw which regions.
}
    \FIXME{AAA @ JED: R2-8 }
	\label{fig:userInterface}

\end{figure}

%% file: tables/Participants.tex
\begin{table}
\centering
\footnotesize
\begin{tabular}{@{}l@{}c@{}c@{}}

\textbf{Academic Discipline} 
& \textbf{Participants}
& ~~\textbf{Gamers}\\
\hline
Agricultural Sciences: 4 unique majors
&    8   
&    2 \\
Business: 3 unique majors
&    5    
&    4 \\
Engineering: 11 unique majors 
&  63   
&  56 \\
Forestry: 3 unique majors 
& 4     
&   4 \\
Science: 10 unique majors 
&  25  
&   20 \\
Liberal Arts: 8 unique majors 
&  9    
&    6 \\
Public Health \& Human Sciences: 2 majors 
&   5     
&     1 \\
Undisclosed 
&     5    
&     4\\
\hline
\textbf{Totals} 
&    124     
&   97   \\
\end{tabular}

\vspace{-5pt}
    \caption{Participant demographics, per academic discipline.}
    \label{table:participants}
\vspace{-10pt}

\end{table}

%% file: figure/RTSexperts.tex
\begin{figure}
	\centering

    \includegraphics[width=.6\linewidth]{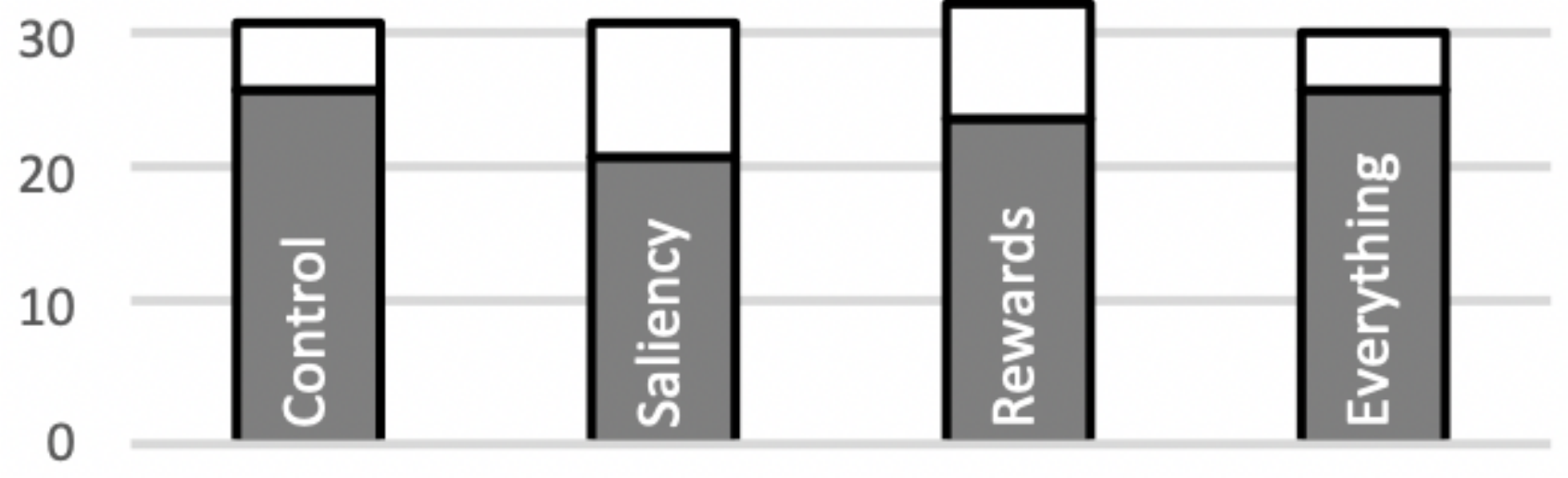}

	\caption{Distribution of RTS ``gamers'' in our study. 
	``Gamers'' are shown in grey, others in white.
	}

	\label{fig:RTSexperts}

\end{figure}

%% file: figure/sequentialDecisions.tex
\begin{figure*}
	\centering
	\includegraphics[width = \linewidth]{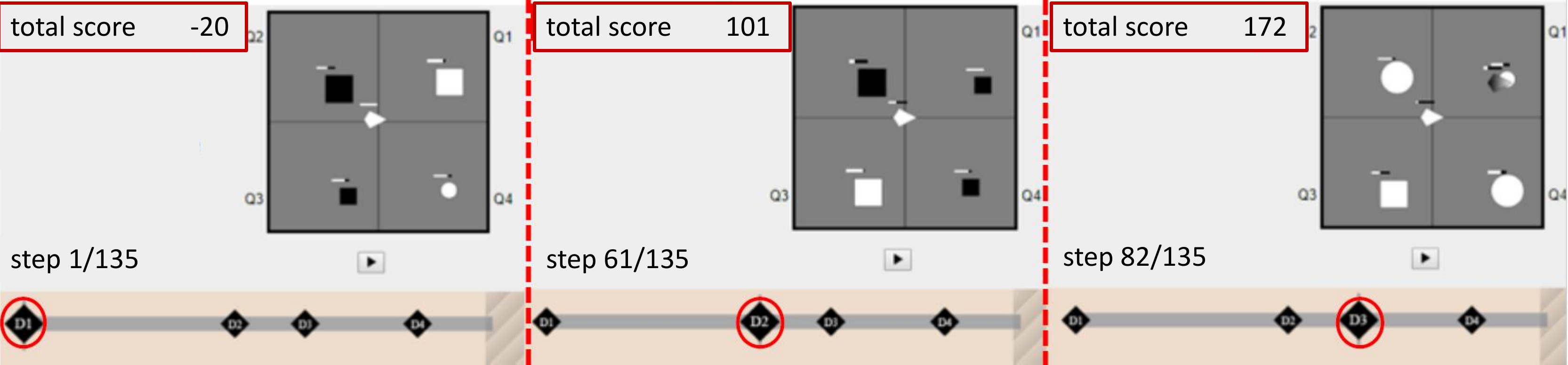}

	\caption{A sequence of the first three DPs of the game.
    For each DP (circled in red) participants saw the game map and the score (boxed in red).
    Next, they made a prediction of which object the AI would choose to attack.
    Last, they would receive an explanation and have the ability to ``play'' the DP.
    At DP1, the agent chose to attack Q2, causing a score change of 121 (+21 pts for damaging and +100 from destroying it).
}
	\label{fig:sequentialDecisions}

\end{figure*}

%% file: figure/GameObjectTable.tex
\begin{figure}
	\centering
    \includegraphics[width=\linewidth]{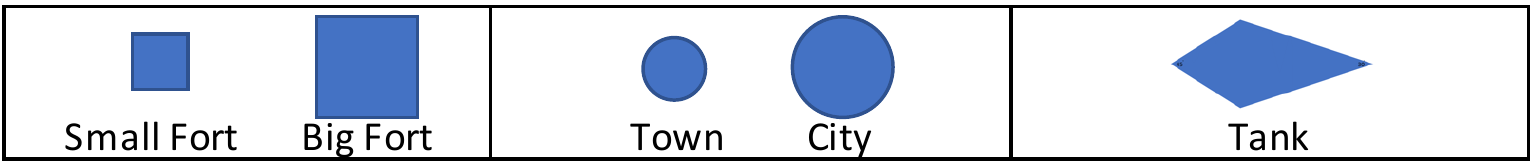}

	\caption{The objects appearing in our game states. Enemy objects were black, and allied objects were white.}

    \label{fig:TableOfObjects}
\end{figure}

%% file: 4-MentalModel.tex
\section{Results: People Describing The AI}\label{AlgorithmDescription}

\boldify{Call back to question we asked participants and why we asked that question}

To elicit participants' understanding of the agent's decision making (RQ-Describe), we used \briefCite{hsieh2005three}'s summative content analysis on participants' answers to the end-of-session question: ``\emph{Please describe the steps in the agent's approach or method...}''~\cite{hoffman2018metrics,lippa2008everyday}.
Figure~\ref{fig:EverythingQuote}
shows a sample response.
\input{figure/EverythingQuote.tex}
 Two researchers independently coded 20\% of the data with 18 codes and reached $> 80\%$ agreement (Jaccard index)~\cite{jaccard1908nouvelles}.
Given this reliability, one researcher did the rest.
\FIXME{JED: need ref Jaccard here (R2-9 here), and nearby we should also clarify our coding method (R3-1)}

In parallel with this process, we generated a rubric of scores to associate with each code.
The codes representing the agent's four basic concepts were each worth 25\% (e.g., its score maximization objective).
Extra nuances in participants' descriptions earned small additions of ``extra credit'' (e.g., saying the agent maximized its \emph{future} score), and extra errors earned small deductions (e.g., saying it tried to preserve its HP). 
Experimenting with different values for the small extras and errors had little effect on comparisons of score distributions among treatments. 
Figure~\ref{fig:MentalModelScore} has the score distribution.

\input{tables/MentalModelCodes.tex}


\subsection{The More, The Better?}\label{sectionMoreBetter}



\boldify{RESULT - The Everything treatment, the most sound and complete treatment, performed significantly better (ANOVA, F=8.369, df=(1,59), p=.00534)}
As Figure~\ref{fig:MentalModelScore} shows, \treatment{Everything} participants had significantly better mental model scores than \treatment{Control} participants
\anova{8.369}{59}{.005\footnote{We consider $p<.05$ significant, and $.05 \leq p <.1$ marginally significant, by convention~\cite{cramer2004sage}.}}.
One possible interpretation is that the \treatment{Everything} participants' performance was due to receiving 
the most sound and complete explanation, consistent with \briefCite{kulesza2015principles}'s results.


\boldify{But are they just relying on one explanation? Let's discuss the rewards}

However, another possibility is that the participants in the \treatment{Everything} treatment were benefiting from only one of the explanation types, and that the other type was making little difference.
Thus, we isolated each explanation type.

To isolate the effect of the reward bars, we compared all participants who saw the decomposed reward bars (the \treatment{Rewards} and \treatment{Everything} treatments) with those who did not.
As the left side of Figure~\ref{fig:combined} illustrates, participants who saw reward bars had significantly better mental model scores than those who did not
\anova{6.454}{122}{.0123}.
Interestingly, isolating the effect of saliency produced a similar impact.
As the right side of Figure~\ref{fig:combined} illustrates, 
those who saw saliency maps (the \treatment{Everything} + \treatment{Saliency} treatments) had somewhat better mental model scores \anova{3.001}{122}{.0858}. 
This suggests that each component brought its own strengths.

\input{figure/MentalModelScore.tex}

\input{figure/combinedGroupMentalModels.tex}


\subsection{Different Explanations, Different Strengths}\label{sectionDiffStrengths}

\boldify{Though the code set was much larger, participants revealed 2 main differences, Score max objective (core curriculum) and the ``extra credit'' codes found in  Table~\ref{table:MentalModelCodes})}

Four of the 18 codes in our mental model codeset revealed nuanced differences among treatments in the participants' understanding of the agent.
Table~\ref{table:MentalModelCodes} lists these four codes.

Participants who saw rewards (\treatment{Rewards} and \treatment{Everything}) often mentioned that the agent was driven by its objective to maximize its score (Table~\ref{table:MentalModelCodes}'s \feature{Maximize Score}). Over 3/4 (36 out of 46) of the people who mentioned this were in treatments that saw rewards. 
For example: 
``\textit{The agent always tried to get as high a possible total sum of all rewards as possible. It valued allies getting damaged in the future as a rather large negative value, and dealing damage and killing enemy forts as rather high positive values.}''
\user{R}{81}{9301203}\footnote{
First letter of participant ID is treatment (\textbf{\underline{C}}ontrol, \textbf{\underline{S}}aliency, \textbf{\underline{R}}ewards, \textbf{\underline{E}}verything).}
and:
``\textit{These costs and rewards are then summed up into an overall cost/reward value, and this value is then used to dictate the agent's action; whichever overall value is greater will be the action that the agent takes.}''
\user{E}{14}{9271202}



\boldify{The two codes, Forward Looking and Paranoia, reveal a more nuanced understanding of the AI's internal reasoning in an algorithmic abstraction}

Some participants who saw rewards also mentioned the nuance that the agent's interest was in its 
\emph{future} score (Table~\ref{table:MentalModelCodes}'s \feature{Forward Looking}), not the present one: 
``\textit{The AI simply takes in mind the unknown of the future rounds and keeps itself in range to be destroyed `quickly' if a future city is under attack...}'' \user{E}{83}{10111806}.
Over 2/3 of the participants (9 out of 13) who pointed out this nuance saw decomposed reward bars.

\boldify{However, the biggest strength of the rewards treatment was the fact that the explanation allowed participants to pick up on the AI's paranoid behavior. In fact, all 8 responses with the paranoia code featured in the treatments that saw the reward bars}
\input{figure/PredictionLineGraph.tex}
%

Even more subtle was the agent's paranoia (Table~\ref{table:MentalModelCodes}'s \feature{Paranoia}).
It had learned Q-value components that reflected a paranoia about receiving negative rewards for attacking its own friendly units. 
Specifically, even though the learned greedy policy appeared to never attack a friendly unit, unless there was no other option, the Q-components for friendly damage were highly negative even for actions that attacked enemies in many cases. 
After investigating, we determined that this was a result of learning via the on-policy SARSA algorithm\footnote{SARSA learns the value of the $\epsilon$-greedy exploration policy, which can randomly attack friendly units. Thus, the learned Q-values reflect those random future negative rewards. However, after learning, exploration stops and friendlies are not randomly attacked.  
}, which learns while it explores.

This paranoia can be a type of ``bug'' in the agents value estimates.
The only 8 participants in the entire study who pointed out this bug were participants who saw rewards.
For example:
``\emph{The AI appears to be afraid of what might happen if a map is generated containing four [friendly] forts or something, in which it can do a lot of damage to itself}.''\user{R}{73}{10031202}.

\boldify{The saliency component of the everything treatment contributed strengths as well, such as it revealing that the AI behaved differently when it was near death (4 control, 3 saliency, 1 rewards, 7 everything)}

On the other hand, participants who saw saliency maps (\treatment{Saliency} + \treatment{Everything}) had a different advantage over the others--noticing how the agent changed behavior when it thought it was going to die (\feature{Episode Over}).
For example, it tended to embark on ``suicide'' missions at the end of a task when its health was low.
About 2/3 (10 out of 15) of the participants who talked about such behaviors were those who saw saliency maps.
As one participant put it:
``\textit{If it cannot take down any structures, it will throw itself to wherever it thinks it will deal the most damage.}''\user{S}{74}{10041805}.


\subsection{Discussion: Which Explanation?}

\boldify{The temptation might be to draw the conclusion that ``We should just give our users everything, since it is sound and complete''. However, this was only one way of measuring how well our participants understood the AI presented to them.}

On the surface, Section~\ref{sectionMoreBetter} suggests that, in explainable systems, the more explanation we give people, the better.
However, Section~\ref{sectionDiffStrengths} suggests that the question of which explanation or combination of explanations is better is more complex -- each type has different strengths, which may matter differently in different situations.
To investigate how situational an explanation type's strength is, we turn next to a qualitative view of how participants fared in individual situations, which we captured with their predictions at each DP.


%% file: figure/EverythingQuote.tex
\begin{figure}
	\centering
    \small
    {\color{black}
        \fbox{\color{black}%
			    \parbox{.97\linewidth}{%
				The agent began worried about damaging its allies\ldots 
                focused little on its own health and made decisions with respect to its allies\ldots
				
				$\quad$ By DP3 it actually 
				\emph{\colorboxBackgroundForegroundText{highlightColor}{black}
				{assigned a positive point value to destroying}
				\colorboxBackgroundForegroundText{highlightColor}{black}
				{itself in the long term because it so heavily weighted potential}
				\colorboxBackgroundForegroundText{highlightColor}{black}
				{damage to allies}}.
				This is because as its health dropped, it would only be able to attack allies in order to stay alive which would cause a massive penalty. 
				
				$\quad$Therefore, the agent decided to  
				\emph{\colorboxBackgroundForegroundText{highlightColor}{black}
				{always attack the largest base}
				\colorboxBackgroundForegroundText{highlightColor}{black}
				{with the most health}}
				so that it would take the most damage which would benefit allies in the long run. \user{E}{23}{10111207}
        		}
            }
        }
    \caption{Top scoring mental model question response.
    The highlighted portions illustrate both ``basic'' and ``extra credit'' concepts, some of which are described in Table~\ref{table:MentalModelCodes}.}
    
    \label{fig:EverythingQuote}
\end{figure}

%% file: tables/MentalModelCodes.tex
\begin{table}
\small
\begin{tabular}{@{}p{0.13\linewidth}c@{}p{0.7\linewidth}@{}}
\textbf{Code} 
& \textbf{Count}~~
& \textbf{Definition}                                                                                                                                             \\ \hline
Maximize Score             
& 46
&\emph{The agent's overall objective is to maximize its long term score.}  \\
\hline
Forward Looking   
& 13
& \emph{The AI looks towards future instances when accounting for the action that it takes now.} \\
\hline
Paranoia            
& 8
& \emph{The AI is paranoid about extending its life too much, expecting penalties when it should not.} \\
\hline
Episode Over          
&15
& \emph{When the AI is nearing death, it behaves differently than it has in previous decision points.}      
\end{tabular}

\vspace{-10pt}
    \caption{The four mental model codes revealing particularly interesting differences in nuances of participants' mental models.}
    \label{table:MentalModelCodes}
\vspace{-10pt}
\end{table}

%% file: figure/MentalModelScore.tex
\begin{figure}
	\centering
    \includegraphics[width = 0.56125\linewidth]{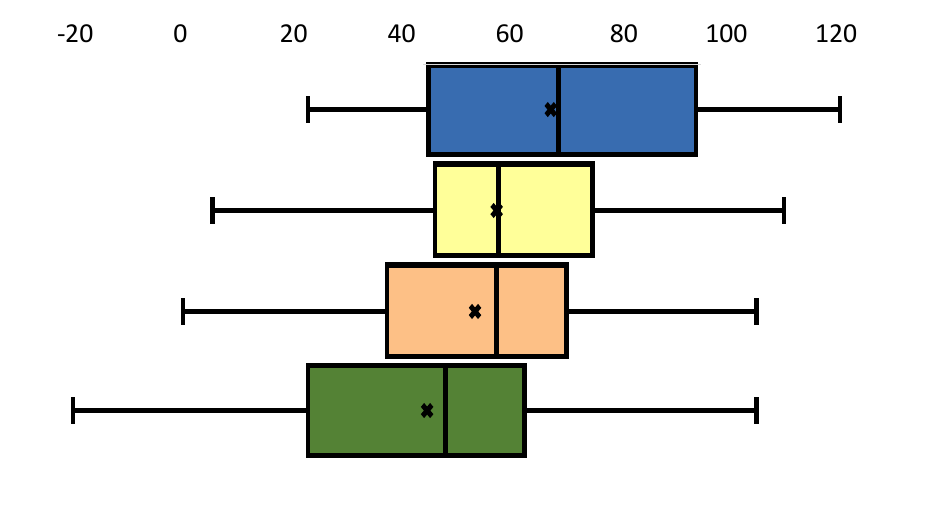}

	\caption{The participants' final mental model scores.
	Box colors from top to bottom:
	    \colorboxBackgroundForegroundText{EverythingColor}{white}{Everything},
        \colorboxBackgroundForegroundText{RewardsColor}{black}{Rewards},
        \colorboxBackgroundForegroundText{SaliencyColor}{black}{Saliency},
        and \colorboxBackgroundForegroundText{ControlColor}{white}{Control}.
	}

    \label{fig:MentalModelScore}

\end{figure}

%% file: figure/combinedGroupMentalModels.tex
\begin{figure}
	\centering
	\includegraphics[width=\linewidth]{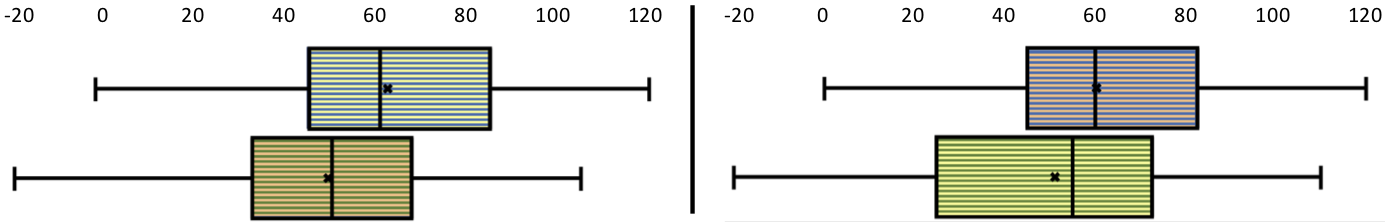}
	\caption{Same data as Figure~\ref{fig:MentalModelScore}.
	Left: Mental model scores for those who saw \emph{rewards} (top) and those who did not.
	Right: Same data, but those who saw \emph{saliency} (top) and those who did not.
	}
	
	
	\label{fig:combined}
\end{figure}


%% file: figure/PredictionLineGraph.tex
\begin{figure*}
	\centering
	\includegraphics[width = \linewidth]{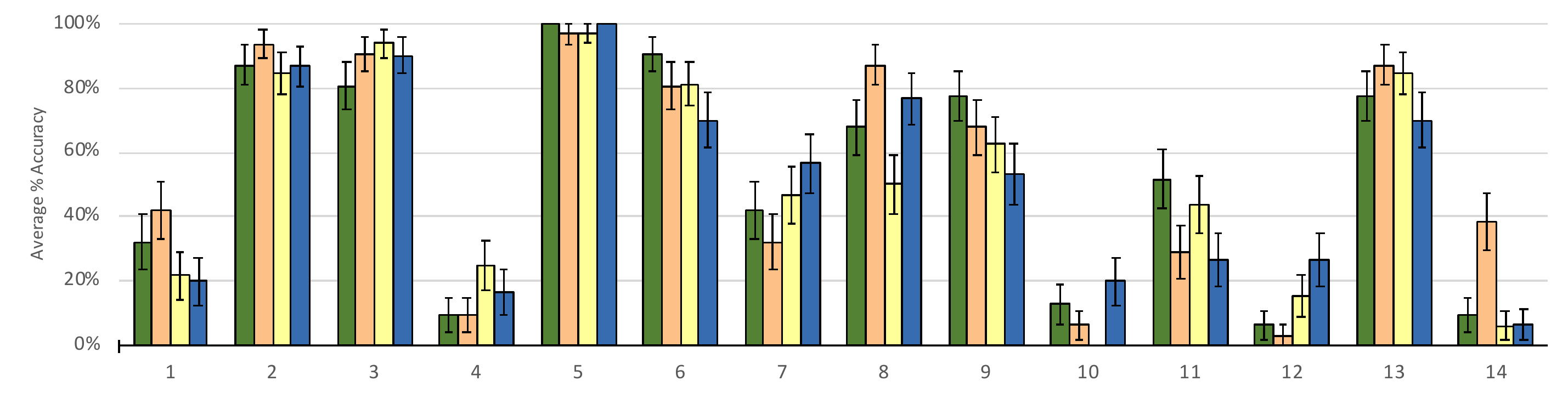}

	\caption{
    Percentage of participants who successfully predicted the AI's next move at each decision point (DP). Bar colors denote treatment (from left to right): 
    \colorboxBackgroundForegroundText{ControlColor}{white}{Control},
    \colorboxBackgroundForegroundText{SaliencyColor}{black}{Saliency},
    \colorboxBackgroundForegroundText{RewardsColor}{black}{Rewards},
    and 
    \colorboxBackgroundForegroundText{EverythingColor}{white}{Everything}. 
    Participants' results varied markedly for the different situations these DPs captured, and there is no evidence that any of the treatments got better over time.
    All error bars ( $SE= \sigma / \sqrt{n} )$ are under 10\%.
    }
	\label{fig:predictiveAccuracy}

\end{figure*}

%% file: 5-PredictionResults.tex
\mysection{Results: People Predicting The AI}\label{resultsPredict}

\boldify{Analysis Methods, rough transplant from methodology.}
\FIXME{
AAA@JED: (R1-2) }

Participants' action predictions  provided us \emph{in situ} data~\cite{hoffman2018metrics,muramatsu2001transparent}.
Recall that at each DP (Table~\ref{table:DecisionPoints}), participants did not see explanations until they predict the agent's action.
As Figure~\ref{fig:predictiveAccuracy} shows, participants' accuracy varied and showed no learning effect.
To understand why, we used qualitative analysis to investigate further, as we detail next.


\input{tables/DecisionPoints.tex}


\mysubsection{Help! The Choice Is Counter-Intuitive}

\boldify{There exists 2 counter-intuitive situations, based on the low participant accuracy}

Situations where the agent went against participants' intuitions proved confusing.
These cases all had low accuracy, with \emph{all} treatments' below random guessing ($\leq$ 25\%).

\boldify{The first flavour: Everything got it right, and when they get it right, they're combining information, but aren't getting better than random guessing.}

One of these situations was the agent choosing neither the strongest nor weakest of similar enemies (DPs 10,12).
When the \treatment{Everything} participants got it right, their comments suggest they combined both saliency and rewards into their reasoning; e.g., \user{E}{71}{10111801} for DP10:
 ``\textit{As it will \textbf{look at the HP} of the tank more it will not attack Q4 instead it will go for Q1 which will \textbf{give it enough benefit} but also maintain its HP}.''

\boldify{When the \treatment{Rewards} treatment got it wrong, they had mixed reasons that indicate they couldn't piece it all together}

However, \emph{all} of the participants in the \treatment{Rewards} treatment got DP10 wrong, suggesting that they needed the saliency maps to factor in how much the AI focuses on its \emph{own} HP, which was key in this situation. 
For example:
``\textit{Lowest HP out of the 3 big fort.}''\user{R}{94}{10031203}. 

\boldify{The last kind of counter-intuitive situations involve those that involve enemy tanks. DP4 is a travesty for predictive accuracy}

A second situation that was counter-intuitive to participants was the agent choosing to attack an enemy elsewhere over saving a friend.
The worst accuracy for this type was at DP4: 77\% of the participants got it wrong.
They incorrectly predicted it would attack the enemy tank, citing its health: ``\textit{This is the enemy object with the lowest value for HP.}''~\user{S}{18}{10121205} or its threat to a friend ``\textit{The enemy tank poses the greatest threat [to] allies...}''\user{S}{25}{9291210}.
Of the few participants (19 total) who predicted correctly, most (68\%) were in  treatments that saw reward bars, e.g.: ``\emph{... destroying enemy [fort] will give you more point than destroying a tank.}''\user{R}{94}{10031203}.

\mysubsection{Overwhelmed!} \label{section:overwhelmed}
\boldify{As discussed before, don't overwhelm your user}



In DPs 6, 9, \& 11, the \treatment{Everything} participants' had the lowest predictive accuracy, while \treatment{Control} had the highest.
This seems tied to \treatment{Everything} participants coping with too much information, showing the importance of balancing completeness without overwhelming users~\cite{kulesza2015principles}. 

Some \treatment{Everything} participants tried to account for \emph{all} the information they had seen.
For example, at DP6: 
``\emph{I think it considers own HP first then Friend/Enemy status, so going by that it will attack Q4. Also, ...it attacks enemies with more HP.}''\user{E}{38}{10131503}.
Some bemoaned the complexity of the information:
``\emph{It was confusing all around to figure out the main factor for movement using the maps and bars...}''\user{E}{39}{9271203}.
In contrast, participants in the \treatment{Control} were able to apply simpler reasoning for the correct Q2 prediction at DP6: ``\emph{because it is the lowest health of all of the enemy objects.}''\user{C}{69}{10011803}.


%

\input{figure/taskTimes}

Participants' timing data also attest to \treatment{Everything} participants' burden of processing all the information (Figure~\ref{fig:taskTimes}).  
In the figure, ``$\times$'' depicts how much time an \treatment{Everything} participant would spend if they spent as long as \treatment{Control}, \emph{plus} the average time  \treatment{Saliency} participants incurred above \treatment{Control}, \emph{plus} the average time  \treatment{Rewards} participants incurred.  
As the figure shows, \treatment{Everything} participants' time to act upon their explanations exceeded the sum of acting upon the component parts, at \emph{every} DP.
Further, since participants had time limits for each DP, some ``timed out''--and \treatment{Everything} accounted for 17 timeouts, while \emph{all others} had 10 in total.





\mysubsection{No Help Needed... Yet}

\boldify{Look at Figure~\ref{fig:predictiveAccuracy} and notice instances like DP 2, 5, and 13: they  were easy for everyone.}

For some DPs (2, 3, 5, 13), explanations seemed unnecessary, as the \treatment{Control} proved ``good enough'' (at least 75\% of participants predicted correctly).
At ``Easy'' situations, explanations may simply interfere.
However, it may not be easy for everyone.
On-demand explanations can provide more information to those who need it, without overwhelming those that do not.


\mysubsection{Discussion: Its All Depends...}

Participants' explanation needs depended on the situation; hence the variability illustrated in Figure~\ref{fig:sequentialDecisions}.
Statistically, treating these situations together simply ``cancels out'' effects.
This wide variation should be expected, given the variability in state/action pairs, combined with the noisiness of \emph{human data}.
The mix of quantitative with qualitative methods for RQ-Predict served us well, and we recommend it to other XAI researchers facing similarly situation-dependent data.



%% file: tables/DecisionPoints.tex
\begin{table}
\begin{tabular}{@{}l|c|c|c|c@{}}
\multirow{2}{*}{\rotatebox{90}{\fontsize{9pt}{10pt} \selectfont Task 1~~~~~}}
                        & {\fontsize{8pt}{12pt} \selectfont DP1} \vspace{-2pt}
                        & {\fontsize{8pt}{12pt} \selectfont DP2}  
                        & {\fontsize{8pt}{12pt} \selectfont DP3}  
                        & {\fontsize{8pt}{12pt} \selectfont DP4}  \\ 
                        & \includegraphics[width=\mapsize]{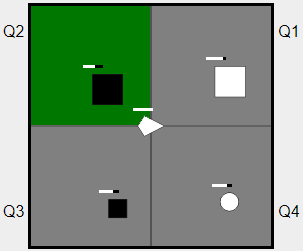}
                        &  \includegraphics[width=\mapsize]{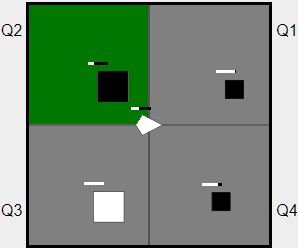}    
                        &  \includegraphics[width=\mapsize]{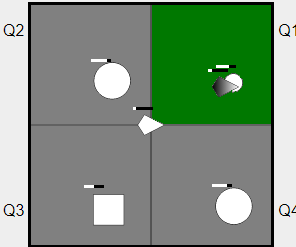}    
                        &  \includegraphics[width=\mapsize]{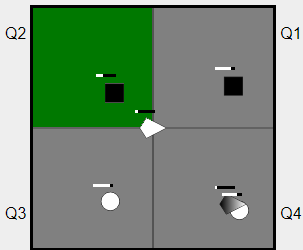}\\ \hline
\multirow{2}{*}{\rotatebox{90}{\fontsize{9pt}{10pt} \selectfont Task 2~~~~~}} 
                   
                        & {\fontsize{8pt}{12pt} \selectfont DP5}  \vspace{-2pt}
                        & {\fontsize{8pt}{12pt} \selectfont DP6}  
                        & {\fontsize{8pt}{12pt} \selectfont DP7}  
                        & {\fontsize{8pt}{12pt} \selectfont DP8} \\ 
                        &  \includegraphics[width=\mapsize]{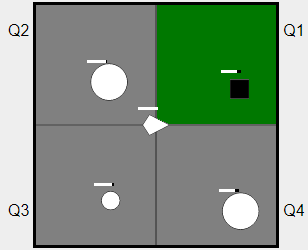}    
                        &   \includegraphics[width=\mapsize]{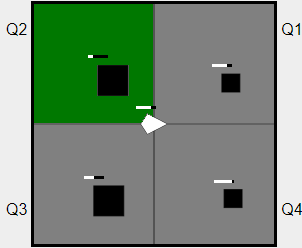}   
                        &  \includegraphics[width=\mapsize]{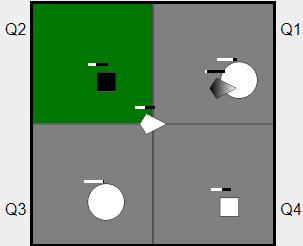}    
                        &   \includegraphics[width=\mapsize]{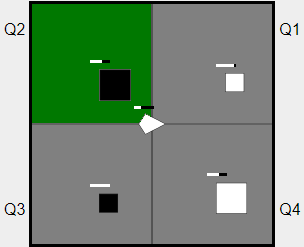}  \\ \hline
\multirow{2}{*}{\rotatebox{90}{\fontsize{9pt}{10pt} \selectfont Task 3~~~~~}} 
                        & {\fontsize{8pt}{12pt} \selectfont DP9}  \vspace{-2pt}
                        & {\fontsize{8pt}{12pt} \selectfont DP10} 
                        & {\fontsize{8pt}{12pt} \selectfont DP11} 
                        &  \\ 
                        &  \includegraphics[width=\mapsize]{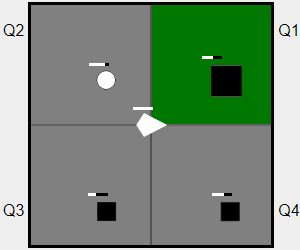}    
                        &   \includegraphics[width=\mapsize]{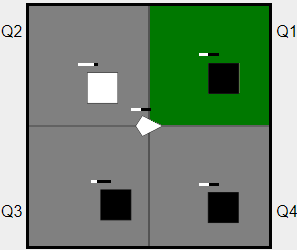}   
                        &   \includegraphics[width=\mapsize]{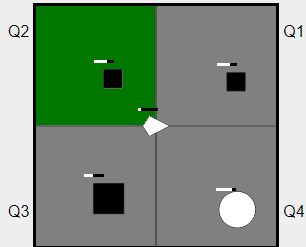}   
                        &    \\ \cline{1-4}
\multirow{2}{*}{\rotatebox{90}{\fontsize{9pt}{10pt} \selectfont Task 4~~~~~}} 
                        & {\fontsize{8pt}{12pt} \selectfont DP12} \vspace{-2pt}
                        & {\fontsize{8pt}{12pt} \selectfont DP13} 
                        & {\fontsize{8pt}{12pt} \selectfont DP14} 
                        &     \\ 
                        &  \includegraphics[width=\mapsize]{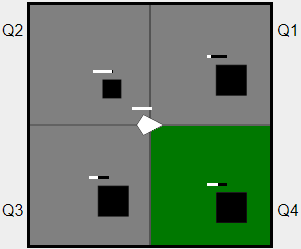}    
                        &   \includegraphics[width=\mapsize]{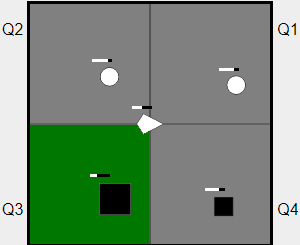}   
                        &   \includegraphics[width=\mapsize]{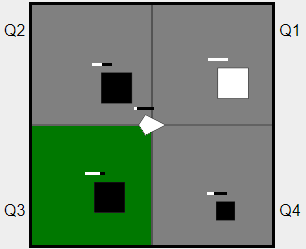}   
                        &   
\end{tabular}

\vspace{-10pt}
    \caption{The tasks and their DPs. We have highlighted in green the action the AI chose.
    }

\vspace{-10pt}
\label{table:DecisionPoints}
\end{table}

%% file: figure/taskTimes.tex
\begin{figure}
	\centering
	\includegraphics[width = \linewidth]{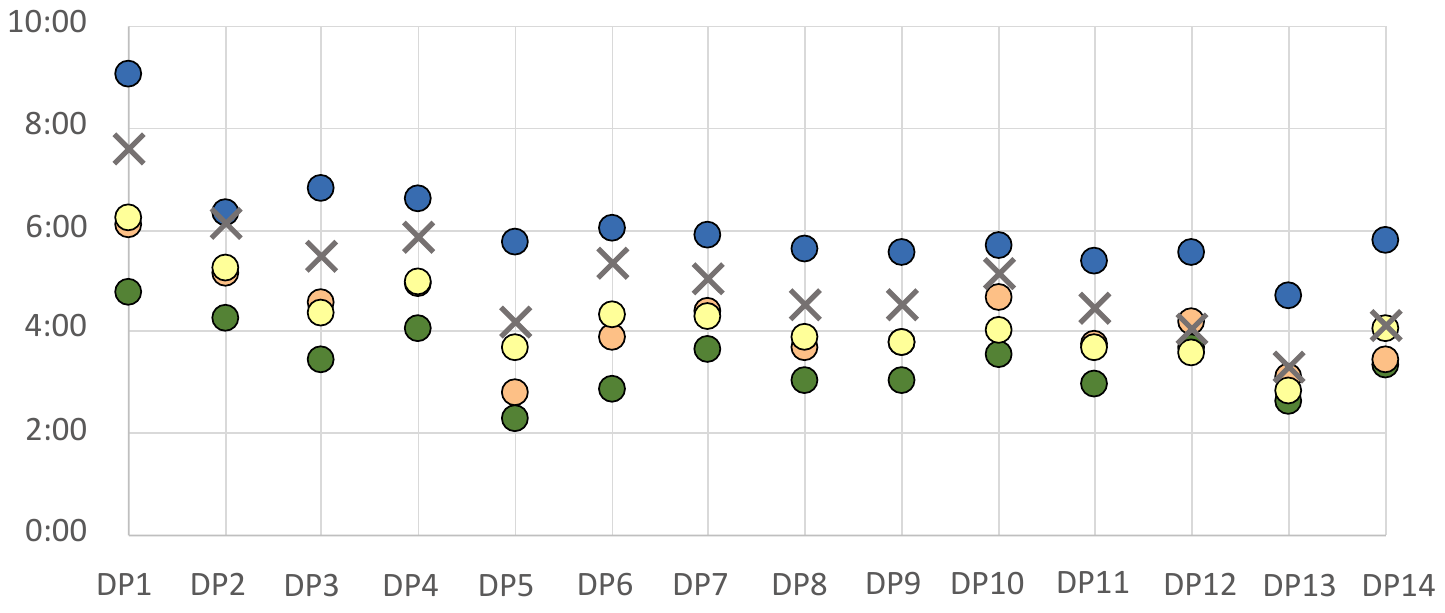}
	
	\caption{Average task time vs DP, per treatment.
	Participants had 12 minutes 
	for the first DP and 8 minutes for all subsequent DPs.
	``$\times$'': see text. Colors: same as Figure~\ref{fig:predictiveAccuracy}
	}
	\label{fig:taskTimes}

\end{figure}

%% file: 7-ThreatsToValidity.tex
\section{Threats To Validity}

\boldify{we did not collect a lot of demographic info, so confounding variables may exist}
Any study has threats to validity~\cite{Wohlin-2012}.
In our study, participants' proficiency in RTS games might have helped understand the agent's tactics.
Although the RTS ``gamers'' were fairly evenly distributed across our treatments (Figure~\ref{fig:RTSexperts}),
we did not collect many demographics, preventing us from considering other factors that may impact people's mental models of games, such as age.
Our study design also emphasized isolation of variables over external validity.
For example, to reduce confounding factors, we simplified our game, but this means that our results might not hold for complex RTS games.
We controlled every participants' time per DP, but this may have impacted their mental models with: insufficient time per DP (8 minutes) and also too few DPs (14).
Threats like these can only be addressed with more studies using diverse empirical methods to generalize the findings.



%

%% file: 8-conclusion.tex
\section{Concluding Remarks}

\boldify{This study  was to see in what ways the explanations help.}

\FIXME{AAA OLD VERSION - 
In this paper, we report on a mixed methods study with 124 participants with no AI background. 
Our goal was to investigate which of four explanation possibilities--saliency, rewards, both, or neither--would enable participants to build the most accurate mental models, and in what circumstances.
Our quantitative results showed that the \treatment{Everything} participants, who saw both saliency and rewards explanations, scored significantly higher on their mental model descriptions than the \treatment{Control} participants.
However, considering the qualitative results in light of the quantitative results for different DPs showed that the type of explanation that helped participants the most was very situation-dependent.

In some situations, full explanation clearly helped, allowing the \treatment{Everything} participants to statistically outperform the \treatment{Control} participants in their descriptions. 
Likewise, Figure~\ref{fig:predictiveAccuracy} shows that \treatment{Everything} participants were top or near-top predictors in about half the DPs. 

Other times, adding explanations caused issues. Some participants in all three of the explanation treatments complained about too much information. 
At almost every DP, \treatment{Everything} participants took longer than the time costs of the ``sum of its parts`` (Figure~\ref{fig:taskTimes}).
Finally, for some DPs, \treatment{Control} participants, who had no explanations, were able to succeed at predicting the AI's move where other treatments' participants failed. 

The results of our quantitative and qualitative analyses suggest several one-size-does-\emph{not}-fit-all takeaway messages.  
First, one type of explanation does not fit all \emph{situations}, as Section~\ref{resultsPredict} shows.
Second, one type of explanation does not fit all \emph{people}, as the distribution ranges in Figure~\ref{fig:MentalModelScore} show.
And perhaps most critical, one type of empirical analysis  (strictly quantitative or strictly qualitative) was not enough; only by combining these techniques were we able to make sense of the wide differences among individual participants at individual DPs. 
We believe that, only by our community applying an arsenal of empirical techniques, can we gain the rich insights needed to learn how to explain AI effectively to mere mortals.
}

In this paper, we report on a mixed methods study with 124 participants with no AI background. 
We investigated which of four visual explanation possibilities--saliency, rewards, both, or neither--enabled participants to build the most accurate mental models, and in what circumstances.
Among the surprising results were:
\begin{itemize}[topsep=0pt,itemsep=0pt,partopsep=0pt, parsep=0pt, leftmargin=*]
    \item \treatment{Everything} participants had significantly better mental model description over the \treatment{Control} participants (Section~\ref{AlgorithmDescription}).
    \item \treatment{Rewards} participants had the most insight into nuanced concepts, such as agent paranoia (Section~\ref{sectionDiffStrengths}).
    \item Participants needed entirely different types of explanations for different situations (Section~\ref{resultsPredict}).
    \item We corroborated \briefCite{kulesza2015principles}'s results about not overwhelming users in a new domain  (Section~\ref{section:overwhelmed}).
\end{itemize}


Our analyses suggest several one-size-does-\emph{not}-fit-all takeaway messages.  
First, one type of explanation does not fit all \emph{situations}, as Section~\ref{resultsPredict} shows.
Second, one type of explanation does not fit all \emph{people}, as the distribution ranges in Figure~\ref{fig:MentalModelScore} show.
And perhaps most critical, one type of empirical analysis  (strictly quantitative or strictly qualitative) was not enough; only by combining these techniques were we able to make sense of the wide differences among individual participants at individual DPs. 
We believe that, only by our community applying an arsenal of empirical techniques, can we gain the rich insights needed to learn how to explain AI effectively to mere mortals.

